# Critical topology and pressure-induced superconductivity in the van der Waals compound AuTe$_2$Br

Erjian Cheng,[1,*] Xianbiao Shi,[2,3,*,‡] Limin Yan,[4,*] Tianheng Huang,[5] Fengliang Liu,[2,6] Wenlong Ma,[7] Zeji Wang,[7] Shuang Jia,[7,8] Jian Sun,[5] Weiwei Zhao,[2,3,9] Wenge Yang,[4,§] Yang Xu,[10,†] and Shiyan Li[1,11,12,¶]

[1] *State Key Laboratory of Surface Physics, and Department of Physics, Fudan University, Shanghai 200438, China*

[2] *State Key Laboratory of Advanced Welding & Joining and Flexible Printed Electronics Technology Center, Harbin Institute of Technology, Shenzhen 518055, China*

[3] *Key Laboratory of Micro-systems and Micro-structures Manufacturing of Ministry of Education, Harbin Institute of Technology, Harbin 150001, China*

[4] *Center for High Pressure Science and Technology Advanced Research, Shanghai 201203, China*

[5] *National Laboratory of Solid State Microstructures, School of Physics and Collaborative Innovation Center of Advanced Microstructures, Nanjing University, Nanjing, 210093, China*

[6] *Department of Physics, Nanchang University, Nanchang, 330031, China*

[7] *ICQM, School of Physics, Peking University, Beijing 100871, China*

[8] *CAS Center for Excellence in Topological Quantum Computation, University of Chinese Academy of Science, Beijing 100190, China*

[9] *Sauvage Laboratory for Smart Materials, School of Materials Science and Engineering, Harbin Institute of Technology, Shenzhen, China*

[10] *Key Laboratory of Polar Materials and Devices (MOE), School of Physics and Electronic Science, East China Normal University, Shanghai 200241, China*

[11] *Collaborative Innovation Center of Advanced Microstructures, Nanjing 210093, China*

[12] *Shanghai Research Center for Quantum Sciences, Shanghai 201315, China*

\* Authors equally contributed to this work.

Corresponding authors: qfxbshi@sina.com (X.B.S.); yangwg@hpstar.ac.cn (W.G.Y.); yxu@phy.ecnu.edu.cn (Y.X.); shiyan_li@fudan.edu.cn (S.Y.L.).

**Abstract**

The study on quantum spin Hall effect and topological insulators formed the prologue to the surge of research activities in topological materials in the past decade. Compared to intricately engineered quantum wells, three-dimensional weak topological insulators provide a natural route to the quantum spin Hall effect, due to the adiabatic connection between them and a stack of quantum spin Hall insulators, and the convenience in exfoliation of samples associated with their van der Waals-type structure. Despite these advantages, both theoretical prediction and experimental identification of weak topological insulators remain scarce. Here, based on first-principles calculations, we show that $AuTe_2Br$ locates at the boundary between a strong and a weak topological insulating state. More interestingly, the critical topology of $AuTe_2Br$ persists up to an applied pressure of ~ 15.4 GPa before a structural phase transition accompanied by a change of electronic topology and the onset of superconductivity. Our results establish $AuTe_2Br$ as a new candidate for weak topological insulators with the potential to realize various other topological phases of matter.

**Introduction**

For the classification of states of matter, the role of topology as a principle beyond spontaneous symmetry breaking was first demonstrated in quantum Hall states[1–9]. However, the extensive current research interest in topological materials started with topological insulators (TIs), preluded by the discovery of quantum spin Hall (QSH) states in quantum wells. QSH insulators, often synonymously referred to as two-dimensional (2D) TIs, are characterized by a finite bulk band gap and gapless helical edge states carrying opposite spins[1–9]. Time-reversal symmetry dictates the destructive interference between backscattering paths, resulting in dissipationless spin currents, bearing huge significance for electronic and spintronic applications[10–13]. Despite the intense pursuit in the past decade for QSH insulators beyond quantum-well systems, the only candidates found are monolayer $WTe_2$ [5–8] and $WSe_2$ [9].

Compared to the case of a time-reversal-invariant 2D TI classified by the $\mathbb{Z}_2$ index, 3D TIs are classified by four $\mathbb{Z}_2$ indices $(v_0; \mathbf{v}) \equiv (v_0; v_1v_2v_3)$. A strong TI (STI) is indicated by a nonzero $v_0$, in which the bulk gap is closed by surface states featuring an odd number of Dirac cones on all surfaces of the sample[1,2]. In contrast, a zero $v_0$ and finite $\mathbf{v}$ point to a weak TI (WTI), in which gapless surface states associated with an even number of Dirac cones appear only on the side surfaces. Compared to 2D TIs and 3D STIs, the territory of WTIs remains relatively less explored, with only a few candidates found to date[13–17], such as $ZrTe_5$, $HfTe_5$, $Bi_{14}Rh_3I_9$, $\beta$-$Bi_4I_4$, and CaSn. All these compounds consist of weakly coupled layers, a structure desirable for realizing WTI, as a WTI is topologically equivalent to a stack of 2D TI layers[13–17]. Considering how the sample can be readily cleaved to few layers or even monolayer, WTIs with a van der Waals-type structure provide a promising platform for realizing QSH states. Moreover, contrary to initial perceptions and what is implied in their name, WTIs have been demonstrated to be robust even under strong disorders, provided the time-reversal symmetry remains intact[13–17].

Here, by performing density functional theory calculations, we found that the van der Waals-type compound $AuTe_2Br$ is an STI or WTI, depending on whether the raw experimental (Expt.) or fully optimized (Opt.) structural parameters are used as input for the calculations, respectively. This dichotomy indicates that $AuTe_2Br$ may locate at the boundary between an STI and a WTI state. Despite its sensitivity to structural variations, the critical topology is found to persist up to an applied pressure of ~ 15.4 GPa, where a pressure-induced structural phase transition from *Cmc*$2_1$ to *Pmmm* occurs. The robustness against pressure might originate from the opposite trend in the pressure dependence of certain structural parameters. The low-pressure TI phase emerges from a topological nodal-line metal (TNLM) protected by mirror symmetry when the spin-orbit coupling (SOC) is ignored. In contrast, the high-pressure phase in the absence of SOC is a TNLM stabilized by a distinct mechanism due to the recovery of inversion symmetry. Furthermore, superconductivity is observed in a wide pressure range in the high-pressure phase. Our results not only add another candidate to the list of WTIs and hence QSH insulators, but also provide a fertile playground for the exploration of topological phase transitions and topological superconductivity.

## Results and discussion

AuTe$_2$X ($X$ = Cl, Br) crystallize in an orthorhombic structure, consisting of halogen atoms inserted between layers of AuTe$_2$ networks stacking along the *b* axis [Fig. 1(a)], despite their space groups differing from each other—*Cmcm* (No. 63) and *Cmc*2$_1$ (No. 36) for Cl, and Br, respectively[18–21]. The topological properties of AuTe$_2$X remained unexplored until recently[21,22]. Transport measurements revealed the compensated semimetal nature of AuTe$_2$Br with ultrahigh carrier mobility of ~ 10$^5$ cm$^2$V$^{-1}$s$^{-1}$ and a nonsaturated magnetoresistance reaching ~ 3 × 10$^5$ at 4.2 K and 58 T [21]. More interestingly, it was proposed that the QSH effect can be realized in monolayer AuTe$_2$Cl through the confinement effect[22]. Our calculations on the band structure of bulk AuTe$_2$Cl (Supplementary Fig. 1) give unanimously a topological nontrivial gap along the Γ–Y direction and a $\mathbb{Z}_2$ of (1; 110), regardless of whether Expt. or Opt. structural parameters are used, confirming the STI nature of AuTe$_2$Cl. Note that such an STI is topologically equivalent to the topological semimetal suggested by previous studies[22], considering the complete separation of the valence and conduction bands.

Our calculation results for AuTe$_2$Br in the monolayer limit are presented in Supplementary Fig. 2. Fully optimized structure parameters are adopted for the monolayer, giving relaxed lattice constants close to the experimental values for the bulk counterpart. Similar to the case of monolayer AuTe$_2$Cl, monolayer AuTe$_2$Br is shown to be a 2D TI, rendering it a candidate platform for the study of the QSH effect.

Turning to the bulk AuTe$_2$Br, the calculation results in the absence of SOC are shown in Figs. 1(c–e) and Figs. 1(f–h) for Expt. and Opt. structural parameters, respectively. The results are similar for the two sets of parameters. We find a metallic ground state with electron pockets at the X and M points and hole pockets at the Γ and Y points. The bands exhibit crossing points in the high symmetry directions Γ-Y and Y-K that are tied to the ΓYKZ plane of the bulk Brillouin zone. By projecting the states onto atomic orbitals, we find that the two bands forming the crossing points are contributed by the

Au $d_{xz+yz}$ orbital and Te $p_z$ orbital, respectively, as shown in Figs. 1(d) and 1(g). The band order is inverted around the Y point, indicating a nontrivial topology. Due to the presence of mirror symmetry associated with the ΓYKZ plane, the crossing points actually form a nodal line in this plane centering around Y, as depicted in Figs. 1(e) and 1(h).

The calculation results for the band structure of bulk AuTe$_2$Br with SOC included are shown in Figs. 2(a–d) and Figs. 2(e–h) for Expt. and Opt. structural parameters, respectively. The nodal line is gapped for both cases. Surprisingly, however, topologically distinct results are found for the two sets of parameters. Before structural optimization, there is a gapped Dirac cone arising from the band hybridizations between Au $d_{xz+yz}$ and Te $p_z$ orbitals along Γ–Z, resulting in inverted bands at the Y point [Fig. 2(b)]. With structural optimization, the band order at Γ is inverted as well [Fig. (f)]. To further characterize the band topology, the Wannier charge centers (WCC) evolution on the time-reversal invariant planes is calculated using the Wilson loop method, as shown in Figs. 2(c) and 2(g). In a WCC plot, the $\mathbb{Z}_2$ indices can be obtained by counting the number of times the WCC curves intersect with any arbitrary horizontal reference line[23]. For example, for the $k_x = 0$ plane, no intersection is found so that the associated $\mathbb{Z}_2 = 0$ [the top left panel in Fig. 2(c)], while for the $k_x = 0.5$ plane, one intersection is found so that the associated $\mathbb{Z}_2 = 1$ [the top right panel in Fig. 2(c)]. For a 3D TI, $v_0$ is determined by the parity of the sum of the $\mathbb{Z}_2$ associated with all the $k_i = 0$ ($i = x, y, z$) and 0.5 planes, whereas $v_i$ ($i = 1, 2, 3$) is determined by the $\mathbb{Z}_2$ associated with the $k_i = 0.5$ ($i = x, y, z$) planes. For the Expt. and Opt. case, respectively, we get (1; 110) and (0; 110) and, thus, an STI and a WTI. This is also reflected in the calculated surface states on the (001) projected surface (note that this is a side surface): one surface Dirac point is observed around $\overline{Y}$ for the Expt. case [Fig. 2(d)], while two Dirac points locate around $\overline{\Gamma}$ and $\overline{Y}$ for the Opt. case [Fig. 2(h)].

The dichotomy between the two cases indicates that AuTe$_2$Br exhibits a critical topology, where the balance is tipped towards an STI or a WTI state by a small variation in the crystal structure. It is therefore interesting to study how the band topology evolves under hydrostatic pressure. To obtain the structural parameters under pressure, high-pressure x-ray diffraction (XRD) measurements were performed, and the results are displayed in Fig. 3(a). With increasing pressure, the $Cmc2_1$ phase persists up to ~ 18.8 GPa. Upon further compression, new diffraction peaks emerge, indicative of a structural phase transition. Enthalpy calculations show that above 15 GPa, the $Pmmm$ symmetry is favored [Figs. 3(b) and 3(c)]. No imaginary modes are found in the phonon calculations above 20 GPa (Supplementary Fig. 4), confirming the robust dynamic stability of this $Pmmm$ phase under high pressure. The lattice constants extracted from Rietveld refinements are displayed in Fig. 3(d), showing a drastic drop of $b$ across the structural transition.

The band structure under pressure before the structural transition is shown in Supplementary Fig. 5. With increasing pressure, the band dispersion along Γ–Y increases. However, for both Expt. and Opt. structural parameters, the band order at Γ and Y is maintained. In other words, the critical topology of $AuTe_2Br$ between an STI and WTI is robust against pressure, provided the space group stays the same.

At first glance, this seems to be inconsistent with the fact that the band topology is sensitive to variations in structural parameters. To solve this apparent discrepancy, we investigate the relation between the critical topology and the structural parameters. As discussed above and schematically shown in Figs. 2(b) and 2(f), the key difference between an STI and WTI state is the band order at Γ. We plot for $AuTe_2Cl$ and $AuTe_2Br$ the dependence of the energy difference of the two bands at Γ, $\Delta E_\Gamma$ [ $\Delta E_\Gamma \equiv E(\text{Au } d_{xz/yz}) - E(\text{Te } p_z)$], on representative structural parameters [Figs. 4(a), 4(c) and 4(e)]. For simplicity, only the data points for ambient pressure are shown. As discussed, $AuTe_2Cl$ being an STI is robust against structural variations. For $AuTe_2Br$, with Expt. structural parameters, the pressure-induced decreasing of the Au-Te bond length $d_{\text{Au-Te}}$ and the distance between Au-Te layers $d_{\text{layer}}$ [Figs. 4(b) and 4(d)] pushes the data points under pressure to the left of the ambient pressure data point Br (Expt.), meaning that $AuTe_2Br$ is always an STI. However, for the Opt. case, one would expect a transition from WTI to STI with the considerable pressure-induced decreasing of $d_{\text{Au-Te}}$ and $d_{\text{layer}}$ since the ambient pressure data point Br (Opt.) is already very close to the boundary. Such a transition is not observed in our calculations (Supplementary Fig. 5). This may be explained by the counteracting effect by the pressure evolution of another parameter, namely, the height of an Au-Te layer $h_{layer}$ [Fig. 1(a)]. As shown in Fig. 4(f), $h_{layer}$ increases with pressure. This means that with increasing pressure, one is traversing to the right of the ambient pressure data point Br (Opt.) in Fig. 4(e), keeping AuTe$_2$Br in the WTI state.

We now turn to the band topology of the high-pressure *Pmmm* phase. In the absence of SOC, the band structure exhibits a 3D metallic nature with some bands featuring large energy dispersion across the Fermi level [Fig. 5(a)]. The 3D Brillouin zone is shown in Fig. 5(b). There are two band crossings along two paths in the $k_x = 0$ plane, Γ–Z and Γ–T [Fig. 5(c)]. The crossings are attributed mainly to the Te–$p_x$ and $p_z$ orbitals. At Γ, the $p_z$ band lies above $p_x$, which is a signature of band inversion, implying the nontrivial band topology of the *Pmmm* phase. Moreover, due to the recovery of the inversion symmetry in this phase, a mechanism different from the ambient pressure one dictates the presence of a nodal line in the $k_x = 0$ plane and centering around Γ [Fig. 5(d)]. Here, the nodal line is protected by the combination of time-reversal, inversion, and spin-rotation symmetries. With the inclusion of SOC, the nodal line is fully gapped with a gap size of ~ 0.02 eV, as shown in Fig. 5(a). The (100) surface band structures are shown in Fig. 5(e). Outside the projection of the nodal line crossings, the surface bands disperse upwards when SOC is ignored. Taking SOC into account, there is a metallic surface band connecting the projected bulk valence and conduction bands. This surface band exhibits spin-momentum-locked spin textures, as displayed in Fig. 5(f).

In addition to the nontrivial band topology, we observed superconductivity in a wide pressure range in the *Pmmm* phase. Temperature-dependent resistance data for AuTe$_2$Br single crystals are shown in Fig. 6(a). A superconducting transition emerges at ~ 19 GPa and persists to the highest pressure of ~51 GPa we measured. We then check the superconducting transition under magnetic field [Fig. 6(b)] and obtain the upper critical field $H_{c2}$ at various pressures [Fig. 6(c)]. The pressure dependence of the superconducting transition temperature $T_c$ is summarized in Fig. 6(d). Nevertheless, the nontrivial band topology of AuTe$_2$Br leaves the possibility of topological superconductivity an interesting topic for future studies.

## Summary

In summary, AuTe$_2$Br is proposed as a new platform for the exploration of WTIs based on our first-principles calculations. Nontrivial topology was found both for the ambient pressure phase and the phase after a pressure-induced structural phase transition, the latter accompanied by superconductivity observed in our measurements. Importantly, our calculations also demonstrate the promising possibility of investigating QSH effect in monolayer AuTe$_2$Br, which can be readily obtained due to the van der Waals-type structure.

The fact that AuTe$_2$Br lies in close proximity to the topological transition between an STI and WTI state holds substantial potential in the continuous tuning of such a transition. Although we have shown that pressure may not be the ideal tuning knob, probably as a result of an opposite trend in the pressure evolution of structural parameters determining the electronic topology, specifically designed variation of these parameters may serve this purpose. This may be achieved by chemical substitution, strain, etc., or a combination with hydrostatic pressure. It would then be more interesting to see how superconductivity is affected when the topological transition is traversed.

## Methods

**Sample synthesis and magneto-resistivity measurements.** AuTe$_2$Br single crystals were synthesized, as described in Ref. 21. The as-grown single crystals are soft, silver flakes with high quality, and the biggest natural plane is (0$l$0) plane [see Supplementary Fig. 3(a)]. Magneto-resistivity measurements under ambient pressure were performed in a physical property measurement system (PPMS; Quantum Design).

**Resistance measurements under pressure.** High-pressure resistance measurements were performed on AuTe$_2$Br single crystals with silicone oil as the pressure transmitting medium (denoted as S1, S2, S3 for different runs) by using a diamond anvil cell (DAC). The experimental pressures were determined by the pressure-induced fluorescence shift of ruby at room temperature before and after each experiment. A direct current four-probe technique was adopted. Resistance measurements were performed with a PPMS.

**XRD measurements under pressure.** AuTe$_2$Br single crystals were ground into powder by using a mortar for use in the high-pressure synchrotron angle dispersive X-ray diffraction measurement. The high-pressure synchrotron XRD experiments were carried out using a symmetric diamond anvil cell (DAC) with a 300-micron culet diamond. A rhenium gasket was drilled by laser with a 90-micron diameter hole as the sample chamber. The sample chamber was filled with a mixture of the sample, a ruby ball, and silicone oil as the pressure transmitting medium. The experimental pressures were determined by the pressure-induced fluorescence shift of ruby. Synchrotron angle-dispersive XRD measurements were carried out at beamline BL15U1 of the Shanghai Synchrotron Radiation Facility (SSRF) using a monochromatic beam of 0.6199 Å. The diffraction patterns were integrated by using the Dioptas software, and Rietveld refinement was performed by using the GSAS software.

**Electronic band structure of AuTe$_2$X (X = Cl, Br).** We carried out first-principles calculations within the framework of the projector augmented wave (PAW) method[24], as implemented in the Vienna Ab initio Simulation Package (VASP)[25,26]. A kinetic energy cutoff of 500 eV and a Γ-centered $k$ mesh of 11 x 11 x 7 were selected in all calculations. The energy and force difference criterion were defined as $10^{-6}$ eV and 0.01 eV/Å for self-consistent convergence. The van der Waals (vdW) corrections[27,28] and spin-orbit coupling (SOC) effect are considered for all calculations. The electronic band structure for AuTe$_2$X (X = Cl, Br) were calculated with experimental structure and fully optimized lattice parameters (PBE+D2). PBE+D2 represents Perdew-Burke-Ernzerhof (PBE)[29], generalized gradient approximation (GGA) functional with D2 empirical vdW corrections. The WANNIER90 package[30–32] was adopted to construct Wannier functions from the first-principles results without an iterative maximal-localization procedure. The WANNIERTOOLS code[33] was used to investigate the topological features of surface state spectra. The high-pressure electronic band structure of AuTe$_2$Br were also calculated with experimental structure (Expt.) and fully optimized lattice parameters (Opt.), respectively. The cell parameters of AuTe$_2$Br under lower pressure for calculations are displayed in Supplementary Table 1.

**High-pressure structure calculations for AuTe$_2$Br.** Beyond the pressure where a pressure-induced structural phase transition takes place, we performed a high-pressure crystal structure searching of AuTe$_2$Br by using an in-house program called MAGUS (machine learning and graph theory assisted universal structure searcher), which is accelerated by the employment of Bayesian optimization and graph theory. This method has been successfully applied in many systems under high pressure, such as compounds inside plants and layered materials[34–37]. The structural optimization and electronic structure calculations were carried out within density functional theory using the projector augmented-wave method as implemented in the VASP. We chose $s^1d^{10}$, $s^2p^4$ and $s^2p^5$ as the valence electrons for Au, Te and Br respectively while using the GGA in the PBE exchange-correlation functional. The plane-wave cutoff was set as 340 eV and the Brillouin zone (BZ) was meshed choosing the gamma-centered Monkhorst-Pack approximately $2\pi \times 0.025$ Å$^{-1}$.

## Acknowledgements

This work was supported by the National Natural Science Foundation of China (Grant Nos. 12174064, 52073075, 12125404, 11974162, U1930401), the Ministry of Science and Technology of China (Grant Nos. 2016YFA0300503), and the Shanghai Municipal Science and Technology Major Project (Grant No. 2019SHZDZX01). S.J. acknowledges the National Key R&D Program of China (2018YFA0305601). J.S. gratefully acknowledges the Fundamental Research Funds for the Central Universities. Part of the calculations were carried out using supercomputers at the High Performance Computing Center of Collaborative Innovation Center of Advanced Microstructures, the high-performance supercomputing center of Nanjing University. W.W.Z. acknowledges the Shenzhen Science and Technology Program (Grant No. KQTD20170809110344233).

Y.X. is sponsored by the Shanghai Pujiang Program (Grant No. 21PJ1403100) and the Natural Science Foundation of Shanghai (Grant Nos. 21JC1402300 and 21ZR1420500).

**Author Contributions**

E.J.C. conceived the idea and designed the experiments. E.J.C. and L.M.Y. were responsible for electrical transport experiments under high pressure. X.B.S. and W.W.Z. performed the electronic band calculations for topology. T.H.H. and J.S. conducted the DFT calculations for high-pressure structure searching. L.M.Y., F.L.L. and W.G.Y. performed XRD measurements under high pressure. S.J. provided single crystals. E.J.C., X.B.S. and Y.X. analyzed the data and wrote the paper. X.B.S., Y.X., W.G.Y. and S.Y.L. supervised the project. E.J.C., X.B.S. and L.M.Y. contributed equally to this work. All authors discussed the results and commented on the manuscript.

**Additional Information:** Correspondence and requests for materials should be addressed to X.B.S. (qfxbshi@sina.com), W.G.Y. (yangwg@hpstar.ac.cn), Y.X. (yxu@phy.ecnu.edu.cn) and S.Y.L. (shiyan_li@fudan.edu.cn).

**Competing interests:** The authors declare no competing interests.

**Data availability:** Source data that support the plots within the paper and other findings of this study are available from the corresponding authors upon reasonable request.

**Captions**

**Figure 1 | Band structure calculations for AuTe$_2$Br without spin-orbit coupling (SOC).** (**a**) Crystal structure of AuTe$_2$Br with the *Cmc*2$_1$ space group. $d_{\text{layer}}$ and $h_{\text{layer}}$ represent the distance between the Au-Te layers, and the height of an Au-Te layer, respectively. (**b**) The 3D bulk Brillouin zone (BZ). (**c**–**e**) are the calculated band structure, orbital projected band structures along the Γ-Y-K directions, and the energy difference map between the Au $d_{xz+yz}$ and Te $p_z$ bands in the ΓYKZ plane of the BZ for AuTe$_2$Br with experimental (Expt.) structural parameters, respectively. (**f**–**h**) are those results with fully optimized (Opt.) structural parameters. The contour of the nodal ring is shown by the red line in (**e**) and (**h**). The background color represents the magnitude of the gap between the Au $d_{xz+yz}$ and Te $p_z$ bands. With the gap increasing, the color changes from white to blue. Red ellipse highlights the zero gap.

**Figure 2 | Band structure calculations for AuTe$_2$Br with SOC included.** (**a**–**d**) are the calculated band structure, orbital projected band structures along the Γ-Y direction, the evolution of the Wannier charge centers, and the calculated surface states on the (001) projected surface for AuTe$_2$Br with Expt. structural parameters, respectively. (**e**–**h**) are those results with Opt. structural parameters.

**Figure 3 | Pressure evolution of the crystal structure of AuTe$_2$Br.** (**a**) X-ray diffraction 2$\theta$ plot of AuTe$_2$Br at room temperature up to 52.2 GPa. The ambient-pressure *Cmc*2$_1$ structure persists to ~ 18.8 GPa, beyond which new diffraction peaks

emerge (each marked with a dashed line and asterisk), indicating the emergence of pressure-induced structural phase transition. (**b**) Crystal structure of the high-pressure AuTe$_2$Br with *Pmmm* space group. (**c**) Enthalpy calculations for several predicted structures. Above 15 GPa, the *Pmmm* symmetry has the lowest enthalpy. (**d**) Pressure dependence of the lattice constants derived from Rietveld refinement.

**Figure 4 | The topology of AuTe$_2$Cl and AuTe$_2$Br as a function of representative structural parameters and pressure.** (**a**), (**c**) and (**e**) show the topology—being a strong topological insulator (STI) or weak topological insulator (WTI)—of AuTe$_2$Cl (labeled as Cl) and AuTe$_2$Br (labeled as Br), as represented by the energy difference between Au-$d_{xz/yz}$ and Te-$p_z$ bands at the Γ point, as a function of Au-Te bond length $d_{\text{Au-Te}}$, the distance between Au-Te layers $d_{\text{layer}}$, and the height of an Au-Te layer $h_{\text{layer}}$, respectively. (**b**), (**d**) and (**f**) show the pressure dependence of $d_{\text{Au-Te}}$, $d_{\text{layer}}$, and $h_{\text{layer}}$, respectively.

**Figure 5 | Band structure calculations for AuTe$_2$Br at 50 GPa.** (**a**) Calculated band structure of AuTe$_2$Br at 50 GPa. The red and blue lines represent the results without and with SOC. (**b**) The 3D bulk Brillouin zone (BZ). (**c**) Orbital-projected band structure near the Γ point. The contour of the nodal ring in the absence of SOC is shown by the red line in (**d**). The background color represents the magnitude of the gap between the Te $p_x$ and Te $p_z$ bands. With the gap increasing, the color changes from white to blue. Red ellipse highlights the zero gap. (**e**) Surface band structures without and with SOC for the (100) surface. (**f**) Fermi surface and corresponding spin texture at a fixed energy $E - E_F = 20$ meV of the topological surface states in the presence of SOC.

**Figure 6 | Pressure-induced superconductivity in AuTe$_2$Br.** (**a**) Temperature dependence of the resistance of AuTe$_2$Br single crystals with silicone oil as the pressure transmitting medium. S1, S2 and S3 represent different runs. (**b**) Magnetic field dependence of the superconducting transition of AuTe$_2$Br single crystal (S2) at 27.8 GPa. The superconducting transition temperatures $T_c$ are defined as shown in the panel. (**c**)

Temperature dependence of the upper critical field $\mu_0 H_{c2}$. Solid lines are a guide to the eye. (**d**) Pressure dependence of $T_c$.

**Figure 1**

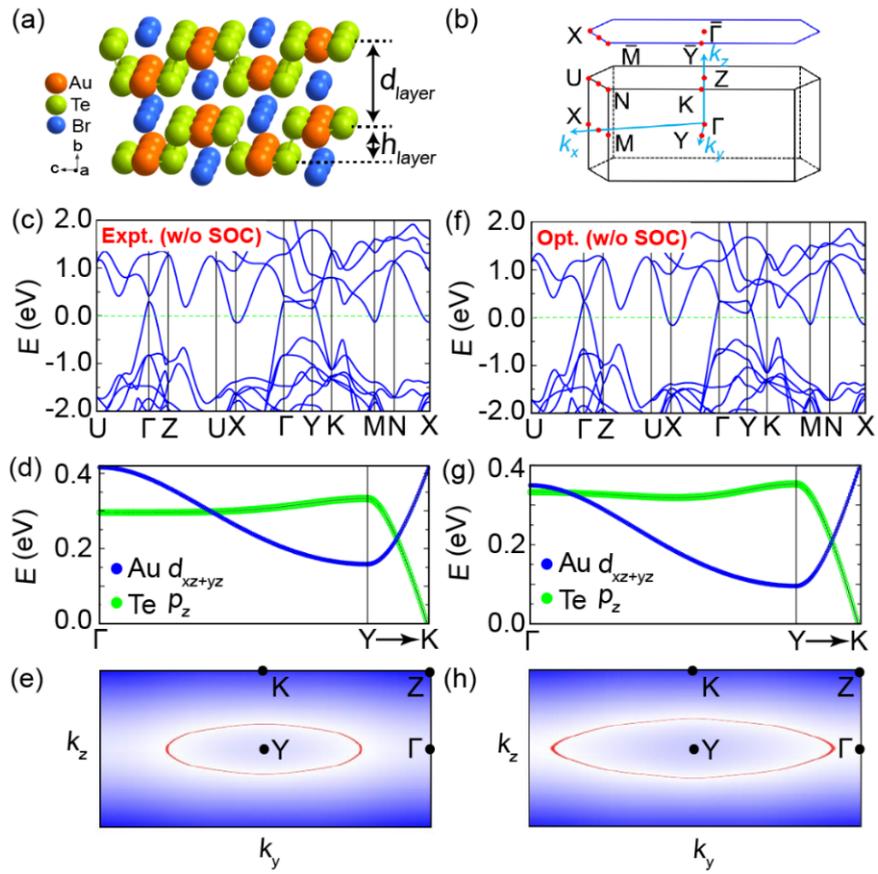

**Figure 2**

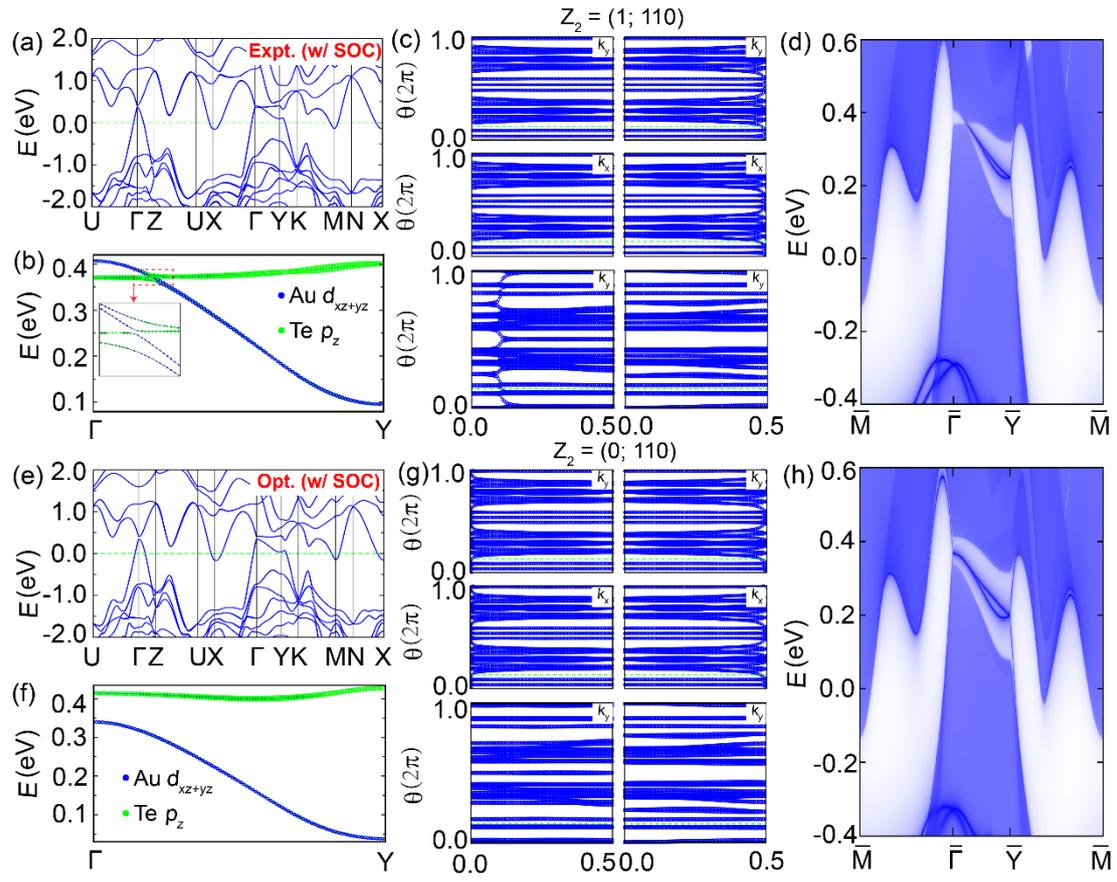

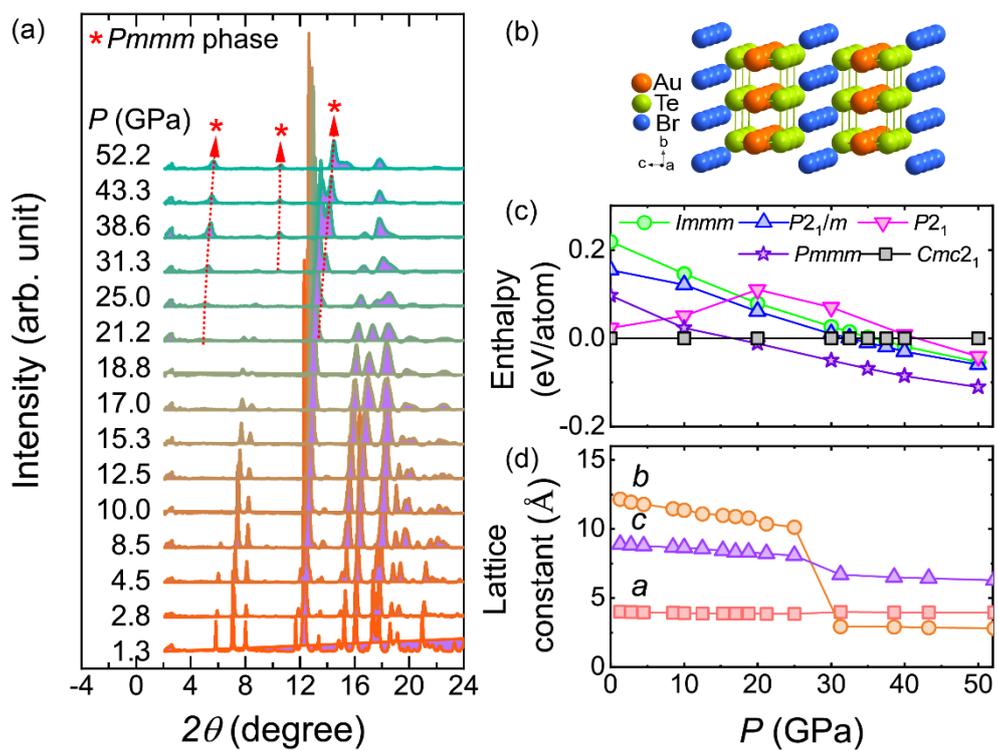

**Figure 4**

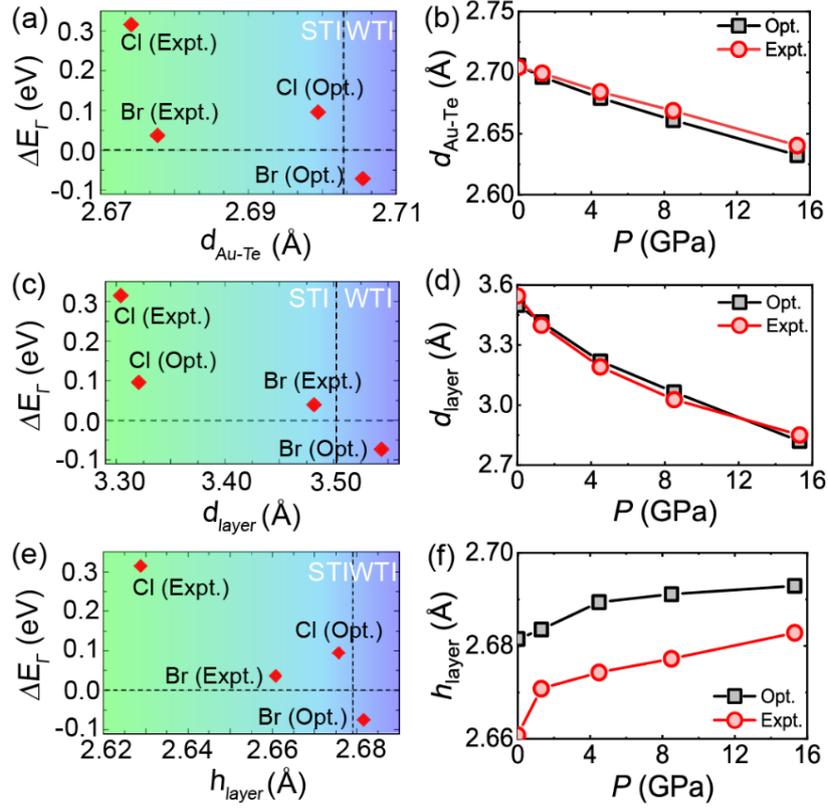

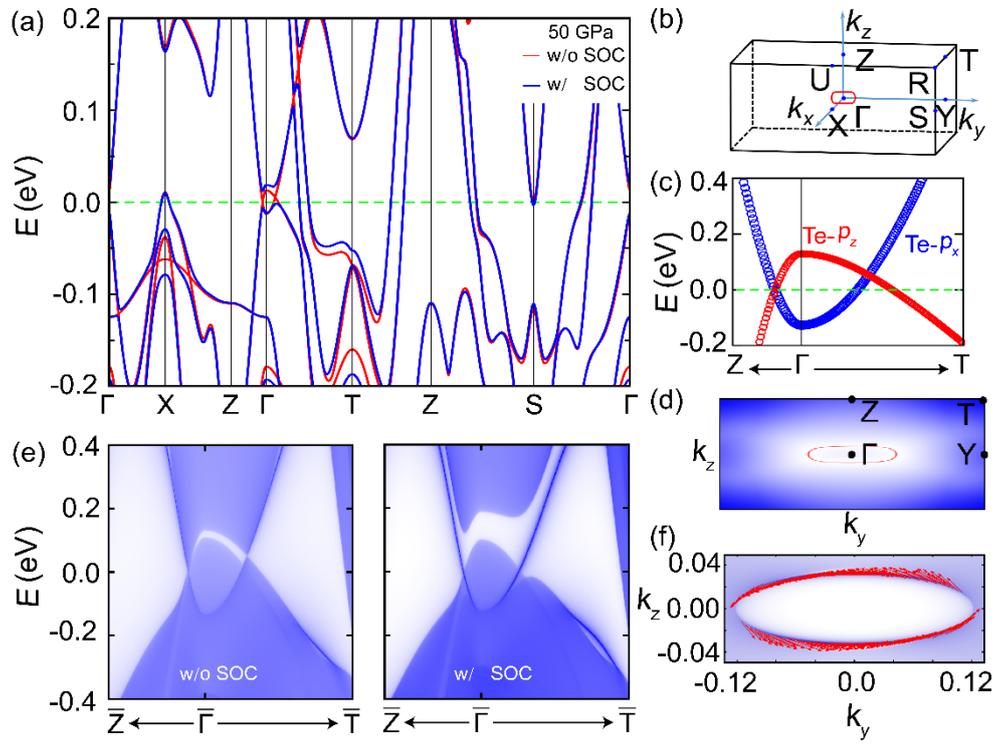

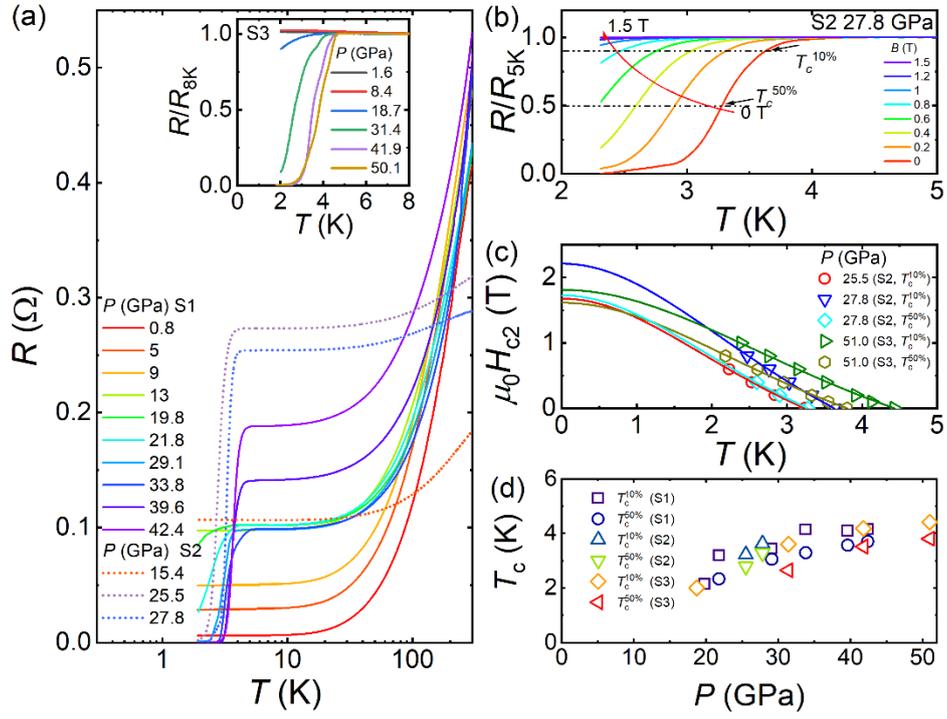

Supplementary Information for

# Critical topology and pressure-induced superconductivity in the van der Waals compound AuTe$_2$Br

Erjian Cheng,[1,*] Xianbiao Shi,[2,3,*,‡] Limin Yan,[4,*] Tianheng Huang,[5] Fengliang Liu,[2,6] Wenlong Ma,[7] Zeji Wang,[7] Shuang Jia,[7,8] Jian Sun,[5] Weiwei Zhao,[2,3,9] Wenge Yang,[4,§], Yang Xu,[10,†] and Shiyan Li[1,11,12,¶]

[1] *State Key Laboratory of Surface Physics, and Department of Physics, Fudan University, Shanghai 200438, China*

[2] *State Key Laboratory of Advanced Welding & Joining and Flexible Printed Electronics Technology Center, Harbin Institute of Technology, Shenzhen 518055, China*

[3] *Key Laboratory of Micro-systems and Micro-structures Manufacturing of Ministry of Education, Harbin Institute of Technology, Harbin 150001, China*

[4] *Center for High Pressure Science and Technology Advanced Research, Shanghai 201203, China*

[5] *National Laboratory of Solid State Microstructures, School of Physics and Collaborative Innovation Center of Advanced Microstructures, Nanjing University, Nanjing, 210093, China*

[6] *Department of Physics, Nanchang University, Nanchang, 330031, China*

[7] *ICQM, School of Physics, Peking University, Beijing 100871, China*

[8] *CAS Center for Excellence in Topological Quantum Computation, University of Chinese Academy of Science, Beijing 100190, China*

[9] *Sauvage Laboratory for Smart Materials, School of Materials Science and Engineering, Harbin Institute of Technology, Shenzhen, China*

[10] *Key Laboratory of Polar Materials and Devices (MOE), School of Physics and Electronic Science, East China Normal University, Shanghai 200241, China*

[11] *Collaborative Innovation Center of Advanced Microstructures, Nanjing 210093, China*

[12] *Shanghai Research Center for Quantum Sciences, Shanghai 201315, China*

\* Authors equally contributed to this work.

Corresponding authors: qfxbshi@sina.com (X.B.S.); yangwg@hpstar.ac.cn (W.G.Y.); yxu@phy.ecnu.edu.cn (Y.X.); shiyan_li@fudan.edu.cn (S.Y.L.).

## Supplementary Note 1: Topological band structure of AuTe$_2$Cl

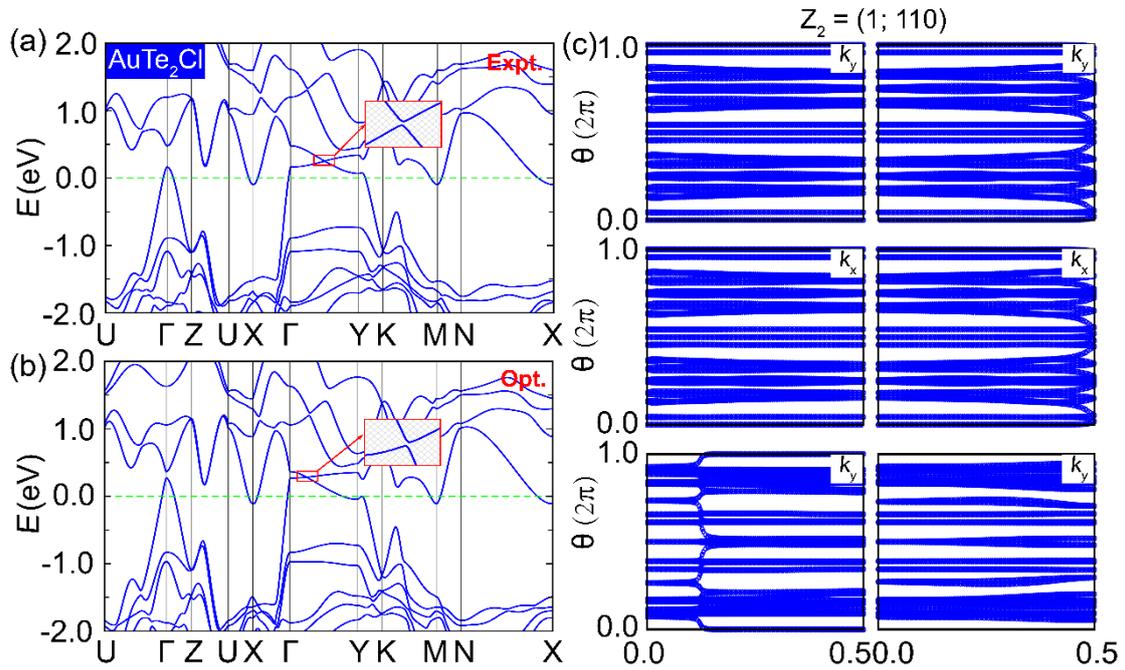

**Supplementary Figure 1 | Band structure of AuTe$_2$Cl.** The calculated band structure for AuTe$_2$Cl with (**a**) experimental structure (Expt.) and (**b**) fully optimized lattice parameters (Opt.). The insets of (**a**) and (**b**) shows the enlarged view of the band dispersion near the Dirac cone. (**c**) The evolution of Wannier charge centers and Z$_2$ topological invariants for AuTe$_2$Cl. These results suggest that AuTe$_2$Cl is a strong topological material.

Supplementary Figs. 1(a) and 1(b) present the band calculations with SOC by using the experimental structure (Expt.) and the fully optimized lattice parameters (Opt.), respectively. For both cases, AuTe$_2$Cl is a strong topological insulator, considering the complete separation of the valence and conduction bands, whose topology is demonstrated by the plot of the evolution of Wannier charge centers and $\mathbb{Z}_2$ topological invariant calculations [Supplementary Fig. 1(c)].

# Supplementary Note 2: Band structure and topology of monolayer AuTe$_2$Br

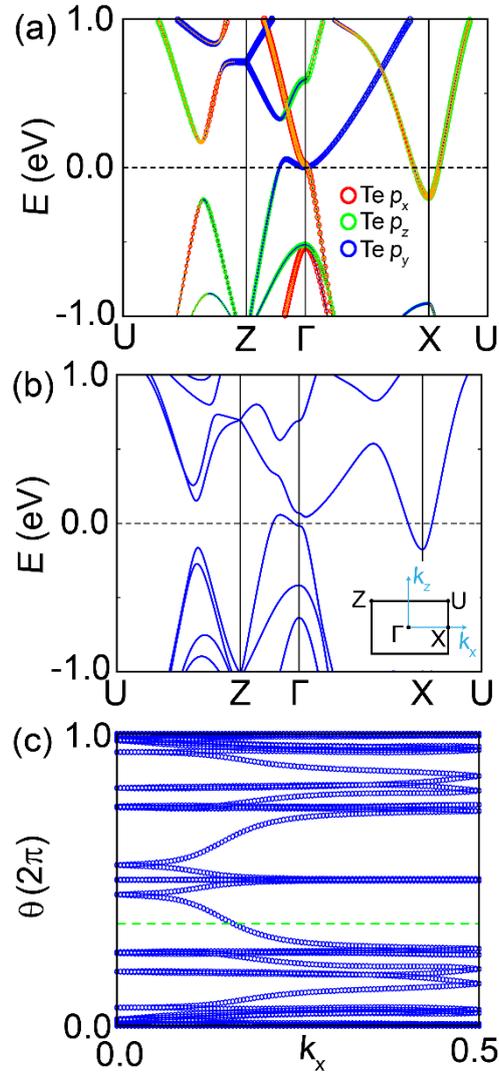

**Supplementary Figure 2 | Band structures of AuTe$_2$Br monolayers.** (**a**) The orbitally resolved band structure of monolayer AuTe$_2$Br without SOC. (**b**) The band structure of monolayer AuTe$_2$Br with SOC. The dashed curve represents the fictitious Fermi level. (**c**) The evolution of WCCs along $k_x$ for monolayer AuTe$_2$Br. The green broken line stands for an arbitrary horizontal line crossing the WCC, and odd times crossing confirms the nontrivial topology.

Supplementary Fig. 2(a) shows the band structure of monolayer AuTe$_2$Br without

SOC. It is found that this system exhibits a metallic state, with two bands crossing the Fermi level. Intriguingly, there is a band crossing point near the Fermi level along the Γ-X direction. By projecting the states onto atomic orbitals, we find that the crossing point along the Γ-X direction is created by a band inversion between the band with Te $p_z$ character and the band contributed by the mixture of Au $d_{xy}$ and Te $p_x$ orbitals at the Γ point. This inverted band structure suggests nontrivial band topology of monolayer AuTe$_2$Br.

With SOC, as shown in Supplementary Fig. 2(b), the crossing point along the Γ-X direction is gaped and there appears a gap between the valence and conduction bands in the whole Brillouin zone. Thus, monolayer AuTe$_2$Br can be considered as an insulator, considering the complete separation of the valence and conduction bands. The $\mathbb{Z}_2$ index is well defined for the bands below the curved Fermi level since there is a finite band gap at each *k* point. It is determined to be 1 by calculating Wannier charge center (WCC), as shown in Supplementary Fig. 2(c), where the number of crossings between an arbitrary horizontal line and the WCC is odd in half of the Brillouin zone.

# Supplementary Note 3: General properties of as-grown AuTe$_2$Br single crystals

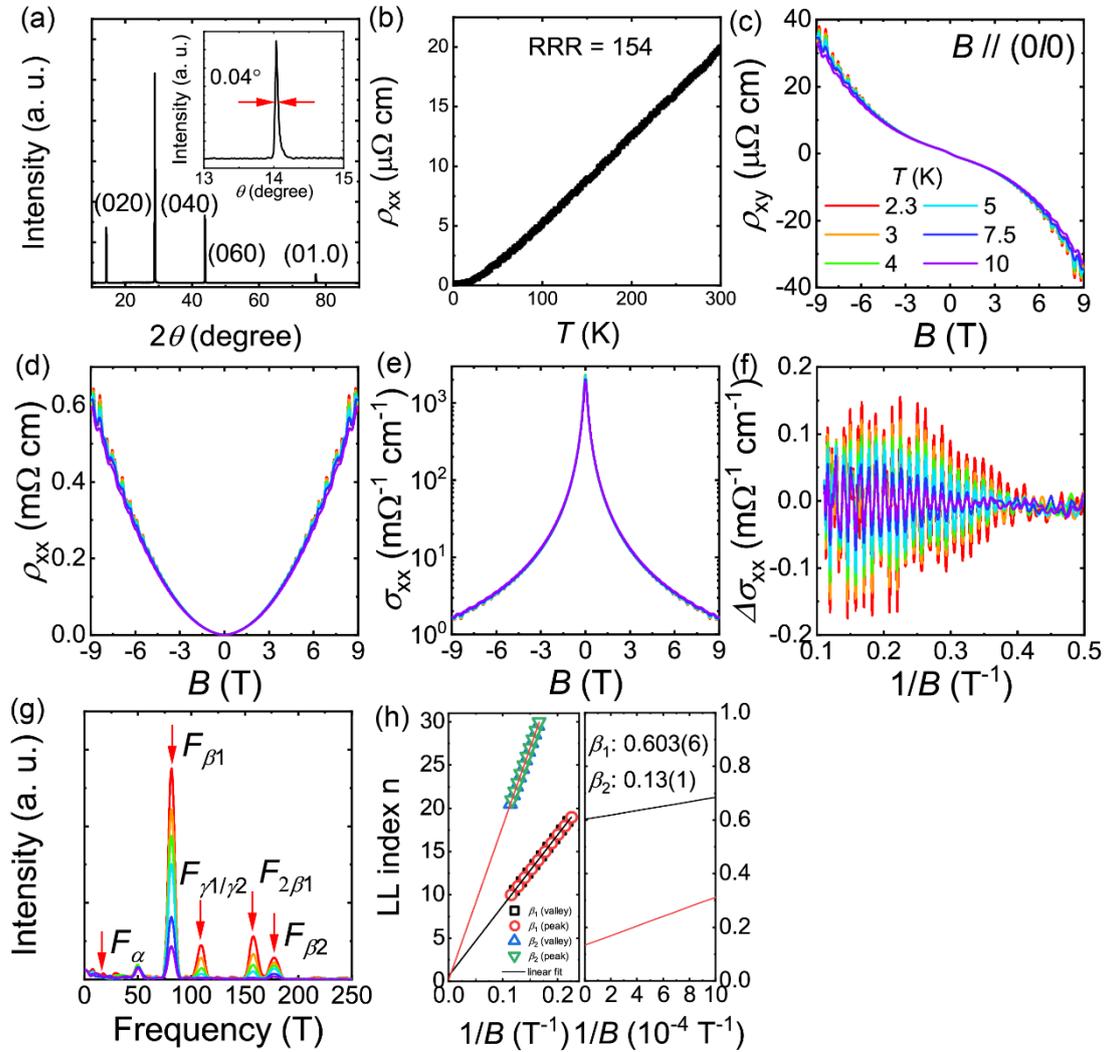

**Supplementary Figure 3 | General properties of as-grown AuTe$_2$Br single crystals.** (**a**) XRD measurements of one piece of single crystal. Inset shows the rocking curve of (040) plane. (**b**) Longitudinal resistivity of a typical single crystal, and the residual resistivity ratio (RRR) is 154. The RRR is defined as $RRR \equiv \rho_{300\,K}/\rho_0$, with residual resistivity $\rho_0 \sim 0.13\,\mu\Omega\,cm$. (**c**) Transverse, (**d**) longitudinal magneto-resistivity and (**e**) longitudinal conductivity at different temperatures. (**f**) The oscillatory component $\Delta\sigma_{xx}$ at different temperatures, extracted from $\sigma_{xx}$ by subtracting a smooth background. (**g**) Fast Fourier transform (FFT) results for Shubonikov-de Haas oscillations (SdH)

oscillations. (**h**) Landau level index plots for the $\beta_1$ and $\beta_2$ bands.

Figure S3 shows the general properties of as-grown $AuTe_2Br$ single crystals with high quality as demonstrated by the small value of the full width at half maximum (FWHM) of 0.04° [the inset to Supplementary Fig. 3(a)], and the large RRR of 154 [Supplementary Fig. 3(b)]. Supplementary Figs. 3(c–h) show the quantum oscillations and the analysis. By analyzing the oscillatory component $\Delta\sigma_{xx}$, we obtain the oscillation frequencies. According to the Landau level index [Supplementary Fig. 3(h)], the intercepts for $\beta_1$ and $\beta_2$ bands locate at 3/8 ~ 5/8 and -1/8 ~ 1/8, respectively, implying $\beta_1$ band has trivial Berry phase whereas $\beta_2$ band has nontrivial Berry phase.

# Supplementary Note 4: Phonon calculations for the high-pressure *Pmmm* phase of AuTe$_2$Br

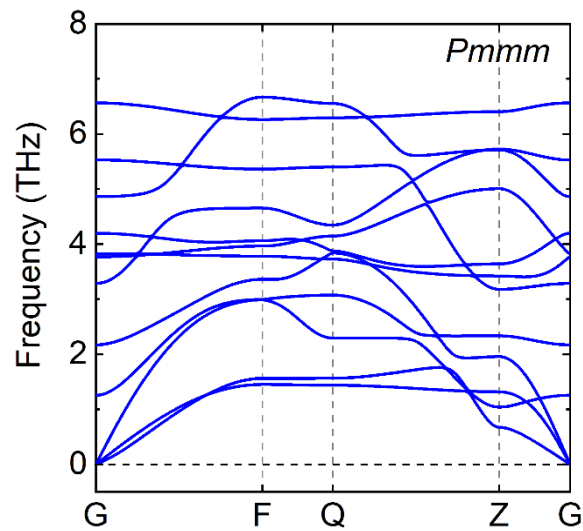

**Supplementary Figure 4 | Phonon calculations for the high-pressure *Pmmm* phase.** No imaginary modes (modes with a negative frequency) are found in the phonon calculations above 20 GPa, confirming the robust dynamic stability of this *Pmmm* phase under high pressure.

# Supplementary Note 5: Crystal parameters of AuTe$_2$Br under pressure

| $P$ (GPa) | | $a$ (Å) | $b$ (Å) | $c$ (Å) |
|---|---|---|---|---|
| 0 | Expt. | 4.018 | 12.284 | 8.885 |
| | PBE | 4.13151 | 13.03588 | 8.93651 |
| | PBE+D2 | 4.02997 | 12.42839 | 8.85382 |
| 1.3 | Expt. | 4.01028 | 12.17008 | 8.8703 |
| | PBE | 4.10543 | 12.69687 | 8.85761 |
| | PBE+D2 | 4.01781 | 12.14308 | 8.76878 |
| 4.5 | Expt. | 3.9685 | 11.7906 | 8.7724 |
| | PBE | 4.0527 | 12.0641 | 8.7285 |
| | PBE+D2 | 3.9951 | 11.7543 | 8.5771 |
| 8.5 | Expt. | 3.9250 | 11.4842 | 8.6754 |
| | PBE | 4.0050 | 11.5730 | 8.5819 |
| | PBE+D2 | 3.9715 | 11.4357 | 8.3828 |
| 15.3 | Expt. | 3.8706 | 11.0043 | 8.4330 |
| | PBE | 3.9396 | 11.1194 | 8.3707 |
| | PBE+D2 | 3.9199 | 11.0878 | 8.1729 |

**Supplementary Table 1 | Crystal parameters of AuTe$_2$Br for band calculations at ambient and high pressures.** The data for experiment (Expt.) is derived from refinement. PBE and PBE+D2 represent Perdew-Burke-Ernzerhof (PBE) GGA functional and PBE GGA functional with D2 empirical VdW corrections, respectively.

## Supplementary Note 6: Band structure of AuTe$_2$Br under pressure

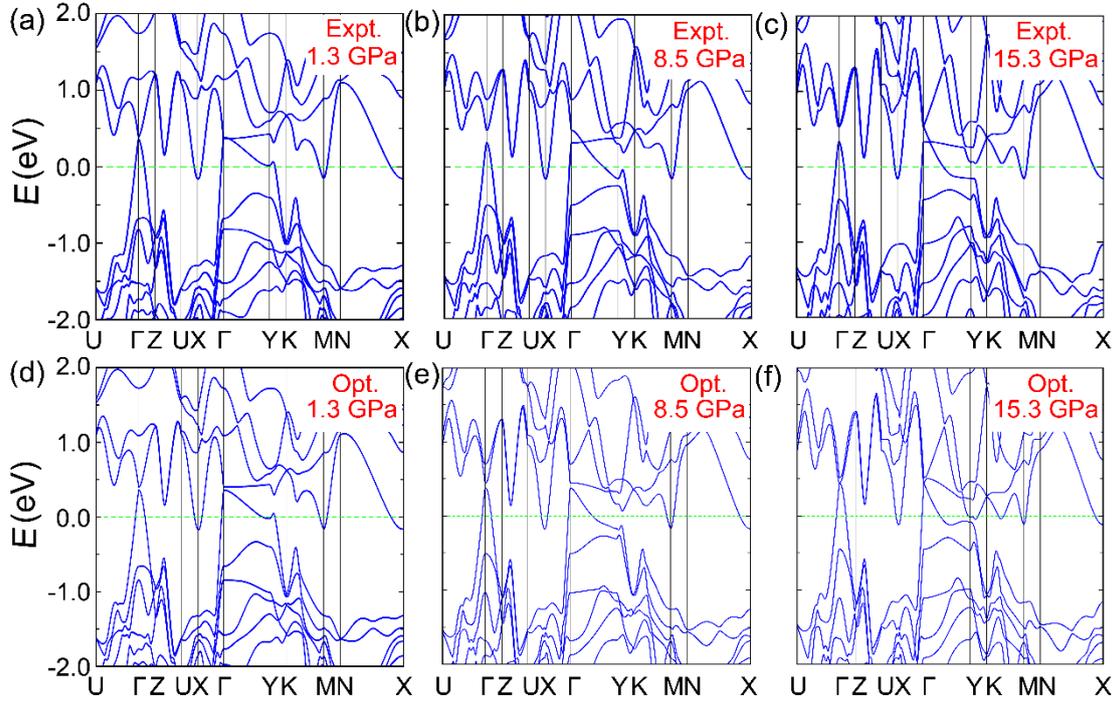

**Supplementary Figure 5 | Electronic band structures of AuTe$_2$Br before structural phase transition by using the experimental data (Expt.) and the fully optimized lattice parameters (Opt.).**

The calculated band structures with experimental input at 1.3, 8.5 and 15.3 GPa are displayed in Supplementary Figs. 5(a–c), respectively, while the calculated band structures by using the fully optimized lattice parameters are shown in Supplementary Figs. 5(d–f), respectively. With increasing pressure, the band structure of AuTe$_2$Br displays more 3D-like feature. We also check the topology behavior at different pressures (data not shown), and found that the topological property of AuTe$_2$Br is robust against pressure, i.e., the band structures with experimental data at different pressures suggest AuTe$_2$Br is a strong topological insulator (STI), while it is a weak topological insulator (WTI) with fully optimized parameters.